\documentclass[%
 preprint,
 amsmath,amssymb,
 aps,
]{revtex4-2}

\usepackage[]{caption}
\usepackage[]{subcaption} 
\usepackage{verbatim}
\usepackage[T1]{fontenc}       
\usepackage{matlab-prettifier}
\usepackage{graphicx}
\usepackage{dcolumn}
\usepackage{bm}
\usepackage{pgfplots}
\usetikzlibrary{arrows,shapes,positioning}
\usepackage{booktabs}

\newenvironment{rcases}
  {\left.\begin{aligned}}
  {\end{aligned}\right\rbrace}

\definecolor{blueSet}{HTML}{1F77B4}
\definecolor{greenSet}{HTML}{2ca02c}
\definecolor{redSet}{HTML}{d62728}

\newlength{\figW}
\newlength{\figH}

\tikzset{
    tikzPlot/.style={trim axis left, trim axis right}
}

\pgfplotsset{
    spaceTimeAxis/.style = {
        width=\figW,
        height=\figH,
        axis x line  = top,
        axis y line  = left,
        xmin         = 0,
        xmax         = \spaceDomainEnd,
        ticks=none,
        y dir = reverse,
        ymin         = 0,
        ymax         = \timeDomainEnd,
        xlabel = {x},
        ylabel = {t},
        ylabel near ticks,
        xlabel near ticks,
        label style={font=\bfseries},
        scale only axis,
    }
}

\begin{document}


\title{On acoustic space-time media that compute their own inverse}

\author{Dirk-Jan van Manen}%
 \email{vdirk@ethz.ch}%
\author{Johannes Aichele}%
\author{Jonas Müller}%
\affiliation{ETH Zurich, Sonneggstrasse 5, 8092 Zurich, Switzerland}%
\author{Marc Serra-Garcia}%
\affiliation{AMOLF, Science Park 104, 1098 XG Amsterdam, The Netherlands}%
\author{Kees Wapenaar}%
\affiliation{TU Delft, Stevinweg 1, 2628 CN Delft, Netherlands}%

\date{\today}

\begin{abstract}
We derive time reflection and transmission coefficients for 1D acoustic waves encountering a time boundary at which the properties of the medium change instantaneously. The time reflection and transmission coefficients are shown to be identical to so-called reverse-space reflection and transmission coefficients which appear in the recursive computation of focusing wavefields used in seismology. We establish a bijectivity between the focusing wavefields and the wavefields produced by time scattering and show how this can be used to construct a space-time medium where the time scattering anticipates the space scattering and ``computes'' the exact inverse for the space scattering. The construction is shown to be independent of the boundary conditions chosen to compute the reflection and transmission coefficients. We demonstrate the construction with a simple numerical example of a single pulse encountering a series of time boundaries before reaching a spatial inhomogeneity. The time boundaries scatter the single pulse into a focusing wavefield that subsequently focuses through the spatial inhomogeneity. Under certain conditions, the transmitted wave has both the same wave shape and amplitude as the original pulse, yielding a transmission coefficient of unity. The reflection coefficient of the space-time medium is always non-zero however.
\end{abstract}

\maketitle

\section{Introduction}
Scattering at time boundaries \citep{morgenthaler, mendonca} offers intriguing new possibilities for wave control. It has recently been exploited to achieve \textit{volumetric} time-reversal \citep{Bacot2016}: Since all waves in a wavefront scatter back energy when ``encountering'' a time boundary, those back-scattered waves exactly retrace the path to the source, no matter how complex the original wavefront was. The distinct properties of time scattering allowed \citet{apffel2022} to achieve frequency conversion of a wave packet. They used a cascade of multiple space and time boundaries to shift the signal over multiple orders of frequency while propagating through the space-time medium. Furthermore, it has been noticed that there is a similarity between time-scattering and reverse-space scattering: \citet{salem2015spacetime} discuss a space-time cross-mapping method to show how time-scattering can be used to facilitate the computation of solutions to certain space-scattering problems. Employing the cross-mapping method, one first computes the inverse of the cross-mapped problem non-recursively using time scattering. The solution to the original problem can then be obtained by inverting the time-scattering solution and inverse cross-mapping it. Here, we go one step further and show how a time medium can be combined with a space medium, to yield a space-time medium that computes its own inverse. Thus, a single pulse entering the  space-time medium on one-side yields a single pulse on the other side and under certain conditions the transmission coefficient of the medium is unity. The reflection coefficient is non-zero however. 

This letter is organized as follows. First, we present the theory for time media with field-continuous boundary conditions. We derive acoustic transmission and reflection coefficients for both time and space boundaries and examine their interrelations. Transmission and reflection coefficients for so-called reverse-space scattering are also derived and motivated. They are shown to be identical to the time transmission and reflection coefficients. The reverse-space coefficients can be used to construct focusing wavefields,  known from seismology \citep{wapenaar_etal_2013, wapenaar_etal_2014}, which are inverses for space scattering. We establish a  bijectivity between the focusing wavefields and the wavefields produced by time scattering. We then explain how a space-time medium can be constructed where the time-scattering anticipates the space-scattering and computes the exact inverse for the space-scattering. Finally, we show that the construction remains valid for time-media with different boundary conditions. The constructions are illustrated with a simple numerical example before discussing and concluding.

\section*{Theory}
Our starting point is the acoustic system of coupled partial differential equations in 1D:
\begin{subequations}
\begin{eqnarray}
    \partial_x p + \rho \partial_t v_x & = & f_x, \\
    \partial_x v_x + \kappa \partial_t p & = & q,
\end{eqnarray}
\label{eq:system_piecewise}
\end{subequations}
where $p$ and $v_x$ are the acoustic pressure and associated horizontal particle velocity, $f_x$ and $q$ are the horizontal volume force and volume injection rate density, $\rho$ and $\kappa$ are the medium mass-density and compressibility, and $\partial_t$ and $\partial_x$ denote partial derivatives with respect to time and space, respectively. In the following, we take $x$ increasing towards the right.

\subsection*{Time transmission and reflection coefficients}
To compute the time transmission and reflection coefficients, we need to specify the boundary conditions.
We start with the acoustic equivalent of taking $\mathbf{E}$ and $\mathbf{H}$ continuous (in the electromagnetic case). Later we will discuss the acoustic equivalent of taking $\mathbf{D}$ and $\mathbf{B}$ continuous. Thus, we take $p$ and $v_x$ to be continuous at the time boundary, i.e.,
\begin{subequations}
\begin{eqnarray}
    p(x, t={0}^{-}) & = & p(x, t={0}^{+}), \\
    v_x(x, t={0}^{-}) & = & v_x(x, t={0}^{+}),
\end{eqnarray}
\label{eq:cc_time}
\end{subequations}
where $t=0$ is the moment that the material properties change, and ${0}^{-}$ and ${0}^{+}$ denote the limit of $t$ approaching zero from negative and positive times, respectively. We define the acoustic impedances in the time media:
\begin{subequations}
\begin{eqnarray}
    \eta_{1} & = & \sqrt{\frac{\rho_1}{\kappa_1}} \quad (t < 0), \\
    \eta_{2} & = & \sqrt{\frac{\rho_2}{\kappa_2}} \quad (t > 0).
\end{eqnarray}
\end{subequations}
Before the time boundary, we consider a wave with amplitude $p_i$ propagating in the positive $x$-direction with propagation velocity $c_1 = (\kappa_1\rho_1)^{-1/2}$:
\begin{equation}
\begin{rcases}
    p & = p_i\delta(x - c_1 t) \\
    v_x & = \eta_1^{-1} p_i\delta(x - c_1 t) 
\end{rcases}
\text{ $t < 0 $ }.
\label{eq:pre_time}
\end{equation}    
After the time boundary, we consider waves with amplitudes~$p_t$ and~$p_r$ propagating in the positive and negative $x$-direction, respectively, with propagation velocity $c_2 = (\kappa_2\rho_2)^{-1/2}$: 
\begin{equation}
\begin{rcases}
    p &= p_t\delta(x - c_2 t) + p_r\delta(x + c_2 t) \\
    v_x &= \eta_2^{-1}\left[p_t\delta(x - c_2 t) - p_r\delta(x + c_2 t)\right] 
\end{rcases}
\text{ $t > 0 $ }.
\label{eq:post_time}
\end{equation}    
Using the continuity conditions, \ref{eq:cc_time}, in combination with the two pairs of equations, \ref{eq:pre_time} and \ref{eq:post_time} above, we obtain:
\begin{subequations}
\begin{eqnarray}
    p_i & = & p_t + p_r, \\
    p_i/\eta_1 & = & p_t/\eta_2 - p_r/\eta_2.
\end{eqnarray}
\end{subequations}
Defining the time transmission and time reflection coefficients, $T^t = p_t/p_i$ and $R^t = p_r/p_i$, respectively, we find:
\begin{subequations}
\begin{eqnarray}
    T^t_{12} & = & \frac{1}{2}\left( 1 + \frac{\eta_2}{\eta_1} \right) =
    \frac{\rho_1c_1+\rho_2c_2}{2\rho_1c_1} ,\label{eq:Tt} \\
    R^t_{12} & = & \frac{1}{2}\left( 1 - \frac{\eta_2}{\eta_1} \right) = 
    \frac{\rho_1c_1-\rho_2c_2}{2\rho_1c_1} 
\end{eqnarray}
\label{eq:timecoeff}
\end{subequations}
where the first digit in the subscript of the transmission and reflection coefficient denotes the incident medium.  

It should be mentioned here, that the boundary conditions~(\ref{eq:cc_time}) imply that the wavenumber spectrum is fixed for waves scattering at a time boundary. Hence, if the propagation velocity changes across the boundary, the scattered waves will experience a frequency shift proportional to the ratio of the propagation velocities~\cite{mendonca}. This will be important when considering under what conditions the space-time medium proposed in the following computes its own inverse.   

\subsection*{Space transmission and reflection coefficients}
We can contrast these Eqs. with the corresponding Eqs. for a spatial boundary at $x=0$ with continuity conditions:
\begin{subequations}
\begin{eqnarray}
    p(x={0}^{-}, t) & = & p(x={0}^{+}, t), \\
    v_x(x={0}^{-}, t) & = & v_x(x={0}^{+}, t),
\end{eqnarray}
\label{eq:cc_space}
\end{subequations}
where ${0}^{-}$ and ${0}^{+}$ are re-defined for the space case as for the time case. Before the spatial boundary, we have an incident and a reflected wave, propagating with amplitudes $p_i$ and $p_r$ in the positive and negative $x$-direction with propagation velocity $c_1$:
\begin{equation}
\begin{rcases}
    p & = p_i\delta(x - c_1 t) + p_r\delta(x + c_1 t)\\
    v_x & = \eta_1^{-1} \left[ p_i\delta(x - c_1 t) - p_r\delta(x + c_1 t)\right]
\end{rcases}
\text{ $x < 0 $. }
\label{eq:pre_space}
\end{equation}    
While after the spatial boundary, we have a transmitted wave, propagating with amplitude $p_t$ in the positive $x$-direction with propagation velocity $c_2$:
\begin{equation}
\begin{rcases}
    p & = p_t\delta(x - c_2 t)\\
    v_x & = \eta_2^{-1} p_t\delta(x - c_2 t)
\end{rcases}
\text{ $x > 0 $. }
\label{eq:post_space}
\end{equation}    
Using the continuity conditions, Eqs.~\ref{eq:cc_space}, in combination with the two pairs of equations,  \ref{eq:pre_space} and \ref{eq:post_space} above, we obtain:
\begin{subequations}
\begin{eqnarray}
    p_i + p_r & = & p_t ,  \\
    p_i/\eta_1 - p_r/\eta_1 & = & p_t/\eta_2 .
\end{eqnarray}
\end{subequations}
Defining the space transmission and reflection coefficients, $T^x = p_t/p_i$ and $R^x = p_r/p_i$, respectively, we find:
\begin{subequations}
\begin{eqnarray}
    T^x_{12} & = & \frac{2\eta_2}{\eta_2+\eta_1} =
    \frac{2\rho_2c_2}{\rho_2c_2+\rho_1c_1} , \label{eq:Tx} \\
    R^x_{12} & = & \frac{\eta_2 - \eta_1}{\eta_2 + \eta_1} = 
    \frac{\rho_2c_2 - \rho_1c_1}{\rho_2c_2 + \rho_1c_1} , \label{eq:Rx}
\end{eqnarray}
\label{eq:spacecoeff}
\end{subequations}
where the first digit in the subscript of the transmission and reflection coefficient denotes the incident medium.  
 
Comparing Eq.~\ref{eq:Tt} for $T^t$, the time transmission coefficient, to Eq.~\ref{eq:Tx} for $T^x$, the space transmission coefficient, we note that the time transmission coefficient is the inverse of the space transmission coefficient, i.e.:
\begin{equation}
    T^t_{12} = \left( T^x_{21}\right) ^{-1}.
    \label{eq:inverted}
\end{equation}

\subsection*{Reverse space transmission and reflection coefficients}
We now analyze scattering at a spatial boundary backwards and derive associated reverse space transmission and reflection coefficients, $T^f$ and $R^f$, respectively.  

We consider the most general case, where waves propagating in the positive and negative $x$-direction are present on both sides of a boundary, denoted by superscripts $^{+}$ and $^{-}$, respectively.
The total $p$ and $v_x$ for $x < 0$ can be written:
\begin{equation}
\begin{rcases}
    p & = p^{+}\delta(x - c_1 t) + p^{-}\delta(x + c_1 t)\\
    v_x & = \eta_1^{-1} \left[ p^{+}\delta(x - c_1 t) - p^{-}\delta(x + c_1 t)\right]
\end{rcases}
\text{ $x < 0 $. }
\label{eq:pre_reverse_space}
\end{equation}    
While for $x > 0$ we have:
\begin{equation}
\begin{rcases}
    p & = p^{+}\delta(x - c_2 t) + p^{-}\delta(x + c_2 t)\\
    v_x & = \eta_2^{-1} \left[ p^{+}\delta(x - c_2 t) - p^{-}\delta(x + c_2 t)\right]
\end{rcases}
\text{ $x > 0 $. }
\label{eq:post_reverse_space}
\end{equation}    
Hence, the total pressure and horizontal particle velocity in a region $j$ can be written in terms of $p^{+}$ and $p^{-}$:
\begin{equation}
    \begin{pmatrix}p \cr v_x\end{pmatrix}_{j} = 
    \begin{pmatrix}1 & 1 \cr \eta^{-1} & -\eta^{-1} \end{pmatrix}_{j}
    \begin{pmatrix}p^{+} \cr p^{-}\end{pmatrix}_{j}.
    \label{eq:compvinc}
\end{equation}
Eq.~\ref{eq:compvinc} is a composition equation. A corresponding decomposition equation is obtained by inverting Eq.~\ref{eq:compvinc}:
\begin{equation}
    \begin{pmatrix}p^{+} \cr p^{-}\end{pmatrix}_{j} = 
    \frac{1}{2}\begin{pmatrix}1 & \eta \cr 1 & -\eta \end{pmatrix}_{j}
    \begin{pmatrix}p \cr v_x\end{pmatrix}_{j}.
    \label{eq:decompvinc}
\end{equation}

Using the fact that $p$ and $v_x$ are continuous across a space interface, we can derive an expression that relates the right- and left-going pressure to the left of an interface (e.g., in layer $1$) in terms of the right- and left-going pressure to the right of an interface (e.g., in layer $2$):
\begin{equation}
    \begin{pmatrix}p^{+} \cr p^{-}\end{pmatrix}_{1} = 
    \frac{1}{2}\begin{pmatrix}1 & \eta \cr 1 & -\eta\end{pmatrix}_{1}
    \begin{pmatrix}1 & 1 \cr \eta^{-1} & -\eta^{-1}\end{pmatrix}_{2}
    \begin{pmatrix}p^{+} \cr p^{-}\end{pmatrix}_{2},
    \nonumber
\end{equation}
or:
\begin{equation}
    \begin{pmatrix}p^{+} \cr p^{-}\end{pmatrix}_{1} = 
    \begin{pmatrix}T^{f}_{21} & R^{f}_{21} \cr R^{f}_{21} & T^{f}_{21}\end{pmatrix}
    \begin{pmatrix}p^{+} \cr p^{-}\end{pmatrix}_{2},
    \label{eq:abovefrombelow2}
\end{equation}
with reverse space transmission and reflection coefficients:
\begin{subequations}
\begin{eqnarray}
    T^{f}_{21} & = & \frac{\eta_2 + \eta_1}{2\eta_2} \quad \textrm{and} \label{eq:Tf}\\
    R^{f}_{21} & = & \frac{\eta_2 - \eta_1}{2\eta_2}, \label{eq:Rf}
\end{eqnarray}
\label{eq:inversecoeff}
\end{subequations}
where the first digit in the subscript of the transmission and reflection coefficient refers to the space medium in the right-hand side of Eq.~\ref{eq:abovefrombelow2}.  

Comparing Eq.~\ref{eq:Tf} with Eq.~\ref{eq:Tx}, we note that these transmission coefficients are inverses of each other:
\begin{equation}
T^f_{21} = \left( T^x_{12}\right) ^{-1}.
\end{equation}
That is: the reverse space transmission coefficient is the inverse of the forward space transmission coefficient.

Furthermore, comparing Eqs.~\ref{eq:inversecoeff} to Eqs.~\ref{eq:timecoeff} for a time material, we see that they have exactly the same form when choosing the material properties of the medium appearing in the right-hand side of Eq.~\ref{eq:abovefrombelow2} as the properties of the incident medium in Eqs.~\ref{eq:timecoeff}:
\begin{subequations}
\begin{eqnarray}
    T^{f}_{21} & = & T^{t}_{21} \quad \textrm{and}\\
    R^{f}_{21} & = & R^{t}_{21} . 
\end{eqnarray}
\end{subequations}
Thus, the time transmission and reflection coefficients are identical to the reverse space transmission and reflection coefficients.

We next explain in detail why it makes sense to identify Eqs.~\ref{eq:inversecoeff} as \textit{transmission} and \textit{reflection} coefficients.

\subsection*{Forward and reverse space scattering}
At first glance, it does not make sense to define the terms appearing in the matrix in Eq.~\ref{eq:abovefrombelow2} as transmission and reflection coefficients. After all, the right-going wave in medium~$1$, $p^{+}_{1}$, is not causally related to the left-going wave in medium~$2$, $p^{-}_{2}$, as such a naming suggests.

To explain the terminology, we contrast two special cases for the right-hand side of Eq.~\ref{eq:abovefrombelow2}. First, we consider the case where the left-going (incident) pressure to the right of the interface vanishes, i.e.:
\begin{equation}
\begin{rcases}
    p^{+}_{1} & = & T^{f}_{21} p^{+}_{2}  \\
    p^{-}_{1} & = & R^{f}_{21} p^{+}_{2}
\end{rcases}
\text{ for $p^{-}_{2} = 0$. }
\label{eq:normal}
\end{equation}    
Second, we consider the case where the right-going (scattered) pressure to the right of the interface vanishes, i.e.:
\begin{equation}
\begin{rcases}
    p^{+}_{1} & = & R^{f}_{21} p^{-}_{2} \\
    p^{-}_{1} & = & T^{f}_{21} p^{-}_{2}
\end{rcases}
\text{ for $p^{+}_{2} = 0$. }
\label{eq:inverse}
\end{equation}    

\begin{figure}
\setlength{\figW}{0.45\columnwidth}
\setlength{\figH}{\figW}

\begin{subfigure}[]{\figW}
 \begin{tikzpicture}[tikzPlot,thick]
        \def\timeBoundaryStart{0.4};
        \def\timeBoundaryEnd{0.6};
        \def\spaceBoundaryStart{0.5};

        \def\spaceDomainEnd{1};
        \def\timeDomainEnd{1};

        \def\cInitial{1};
        \def\cBoundry{1.2};

        \begin{axis}[spaceTimeAxis]
            \draw[thick] (axis cs:0.5,0.1) -- (axis cs:0.5,0.9);

            \draw[thick,-{latex},blueSet, shorten >= 4pt] (axis cs:0.1,0.1) -- node [below left] {$p^+_1$} (axis cs:0.5,0.5);
            \draw[thick,-{latex},blueSet, shorten <= 4pt] (axis cs:0.5,0.5) -- node [above right] {$p^+_2$} (axis cs:0.9,0.9);
            \draw[thick,-{latex},greenSet, shorten <= 4pt] (axis cs:0.5,0.5) -- node [above left] {$p^-_1$} (axis cs:0.1,0.9);
                
            \node [] at (axis cs:0.3,0.95) {$\rho_1,c_1$};
            \node [] at (axis cs:0.7,0.95) {$\rho_2,c_2$};

        \end{axis}
    \end{tikzpicture}
    \caption{forward}
 \end{subfigure}\hfill
 ~
 \begin{subfigure}[]{\figW}
 \begin{tikzpicture}[tikzPlot,thick]
        \def\timeBoundaryStart{0.4};
        \def\timeBoundaryEnd{0.6};
        \def\spaceBoundaryStart{0.5};

        \def\spaceDomainEnd{1};
        \def\timeDomainEnd{1};

        \def\cInitial{1};
        \def\cBoundry{1.2};

        \begin{axis}[spaceTimeAxis]

            \draw[thick] (axis cs:0.5,0.1) -- (axis cs:0.5,0.9);

            \draw[thick,-{latex},blueSet, shorten >= 4pt] (axis cs:0.1,0.1) -- node [below left] {$p^+_1$} (axis cs:0.5,0.5);
            \draw[thick,-{latex},greenSet, shorten >= 4pt] (axis cs:0.9,0.1) -- node [below right] {$p^-_2$} (axis cs:0.5,0.5);
            \draw[thick,-{latex},greenSet, shorten <= 4pt] (axis cs:0.5,0.5) -- node [above left] {$p^-_1$} (axis cs:0.1,0.9);
                
            \node [] at (axis cs:0.3,0.95) {$\rho_1,c_1$};
            \node [] at (axis cs:0.7,0.95) {$\rho_2,c_2$};

        \end{axis}
    \end{tikzpicture}
 \caption{backward}
 \end{subfigure}
  \caption{Space scattering.}
  \label{fig:space_scattering}
\end{figure}
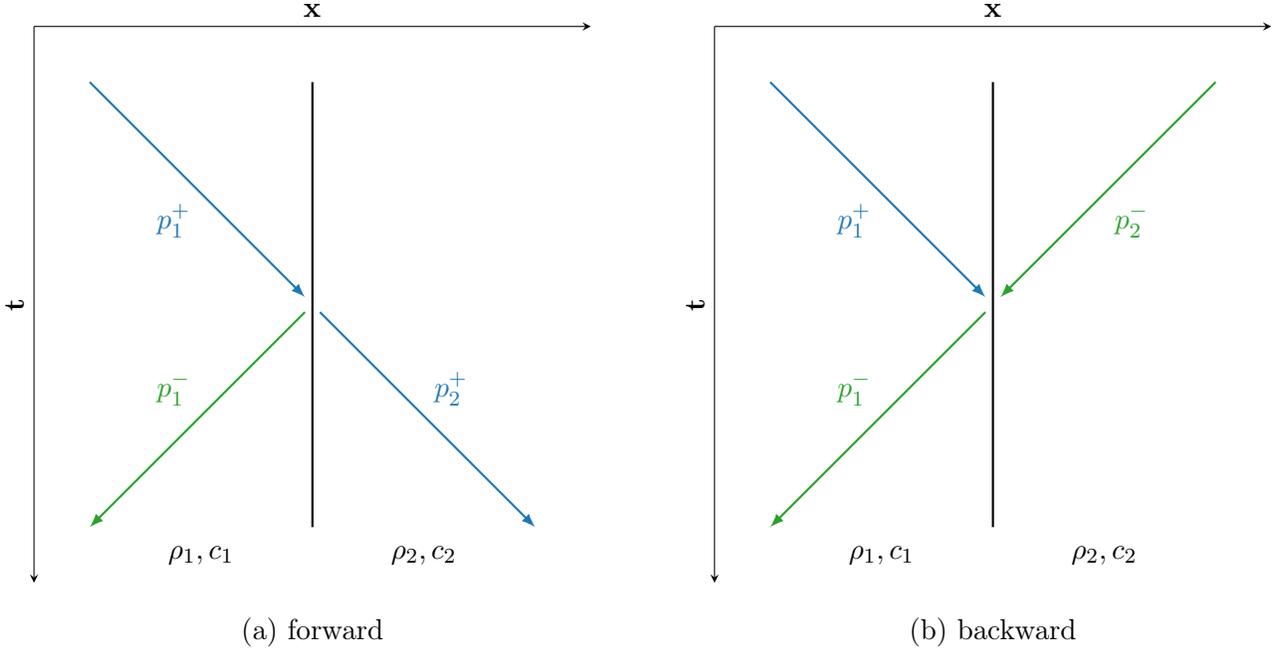

The interpretation of these two cases is as follows. Starting with Eq.~\ref{eq:normal} and Fig.~\ref{fig:space_scattering}, left, the involved fields are the same as those occurring in \textit{normal} forward scattering, namely, a wave $p^{+}_{1}$ incident on an interface from the left, and subsequent reflection and transmission at and through the interface, respectively. However, since we are computing the amplitude of the right- and left-going waves on the left from the amplitude of the right-going wave on the right of the interface, it is clear that we are considering the normal scattering process in reverse. 

On the other hand, Eq.~\ref{eq:inverse} and Fig.~\ref{fig:space_scattering}, right, the involved fields are different from those occurring in \textit{normal} forward scattering since, normally, both the transmission of the incident wave $p^{+}_{1}$ from the left, as well as the reflection of the incident wave $p^{-}_{2}$ from the right, produce scattered, right-going waves $p^{+}_{2}$. The only way for these to cancel each other is if the transmitted wave from the left destructively interferes with the reflected wave from the right. That means that there must be a very specific amplitude and phase relation between the incident waves from the left and from the right. That is exactly what Eq.~\ref{eq:inverse} allows us to compute. 

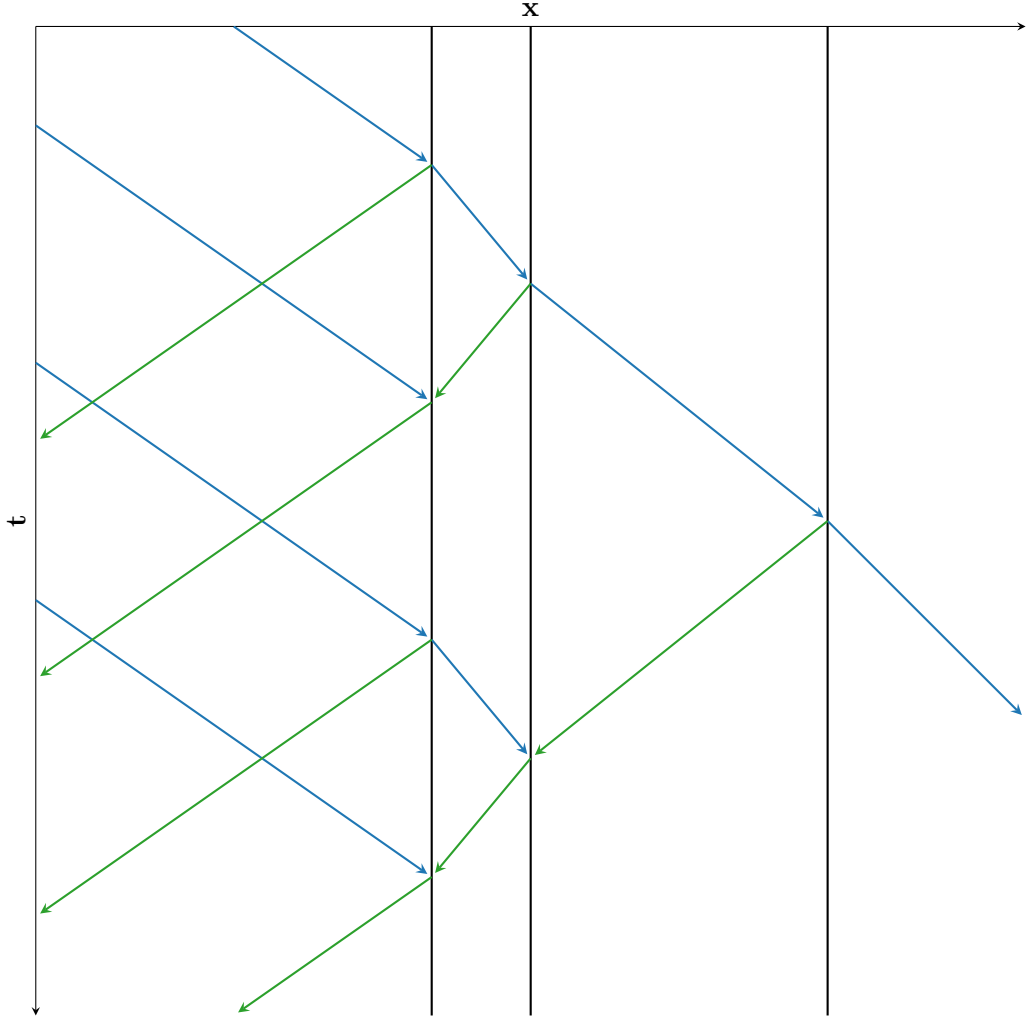
\begin{figure}
\setlength{\figW}{0.8\columnwidth}
\setlength{\figH}{\figW}
 
 \begin{tikzpicture}[tikzPlot]
    \def\timeBoundaryStart{0.4};
    \def\timeBoundaryEnd{0.6};
    \def\spaceBoundaryStart{0.5};
    
    \def\spaceDomainEnd{1};
    \def\timeDomainEnd{1};
    
    \def\cInitial{1};
    \def\cBoundry{1.2};
    
    \begin{axis}[spaceTimeAxis]
        
        \addplot[black, quiver={u=\thisrow{dx},v=\thisrow{dt}},thick,-] table [x={x}, y={t}]{spaceBoundaries.dat};
        
        \addplot[blueSet,quiver={u=\thisrow{dx},v=\thisrow{dt}},thick,-{stealth[sep= 2pt]}] table [x={x}, y={t}]{raysFocusing.dat};
        \addplot[greenSet,quiver={u=\thisrow{dx},v=\thisrow{dt}},thick,-{stealth[sep= 2pt]}] table [x={x}, y={t}]{raysDump.dat};
        
    \end{axis}
\end{tikzpicture}
  \caption{Focusing wavefield by recursive construction.}
  \label{fig:focusing_function}
\end{figure}

\subsection*{Focusing wavefields}
Because the two special cases, Eqs.~\ref{eq:normal} and~\ref{eq:inverse} (Fig.~\ref{fig:space_scattering}), each only consider a single right-going, respectively, left-going wave as input, they can be used to construct a wavefield recursively, without leading to an infinite recursion (see Fig.~\ref{fig:focusing_function}). For example, starting with a single, right-going wave, $p^{+}_{j+1}$, just right of an interface separating layers $j$ and $j+1$, application of Eqs.~\ref{eq:normal} yields the right- and left-going waves $p^{+}_{j}$ and $p^{-}_{j}$ just to the left of that interface. These right- and left-going waves can then be extrapolated further leftwards, to just right of the interface separating layers $j-1$ and $j$, by applying appropriate phase factors. Leftwards extrapolation of a \textit{right-going} wave amounts to propagating the wave \textit{backward} in time and therefore the phase factor should \textit{remove} phase. Leftwards extrapolation of a \textit{left-going} wave amounts to propagating the wave \textit{forward} in time and therefore the phase factor should \textit{add} phase. Eqs.~\ref{eq:normal} and~\ref{eq:inverse} can then be applied again to these individual, leftward-extrapolated waves, yielding two pairs of right- and left-going waves (i.e., four events in total) to be extrapolated. The whole process is repeated, until the leftmost interface is crossed, to yield the complete focusing wavefield (see Fig.~\ref{fig:focusing_function}).

\subsection*{Time-scattering}
\begin{figure}
\setlength{\figW}{0.45\columnwidth}
\setlength{\figH}{\figW}

\begin{subfigure}[]{\figW}
 \begin{tikzpicture}[tikzPlot,thick]
        \def\timeBoundaryStart{0.4};
        \def\timeBoundaryEnd{0.6};
        \def\spaceBoundaryStart{0.5};

        \def\spaceDomainEnd{1};
        \def\timeDomainEnd{1};

        \def\cInitial{1};
        \def\cBoundry{1.2};

        \begin{axis}[spaceTimeAxis]
            \draw[thick] (axis cs:0.1,0.5) -- (axis cs:0.9,0.5);

            \draw[thick,-{latex},greenSet, shorten <= 4pt] (axis cs:0.5,0.5) -- node [below right] {$p$} (axis cs:0.1,0.9);
            \draw[thick,-{latex},blueSet, shorten <= 4pt] (axis cs:0.5,0.5) -- node [below left] {$p$} (axis cs:0.9,0.9);

            \draw[thick,-{latex},greenSet, shorten >= 4pt] (axis cs:0.9,0.1) -- node [above left] {$p$} (axis cs:0.5,0.5);
                
            \node [] at (axis cs:0.9,0.4) {$\rho_1,c_1$};
            \node [] at (axis cs:0.9,0.6) {$\rho_2,c_2$};

        \end{axis}
    \end{tikzpicture}
    \caption{leftgoing}
 \end{subfigure}\hfill
 ~
 \begin{subfigure}[]{\figW}
 \begin{tikzpicture}[tikzPlot,thick]
        \def\timeBoundaryStart{0.4};
        \def\timeBoundaryEnd{0.6};
        \def\spaceBoundaryStart{0.5};

        \def\spaceDomainEnd{1};
        \def\timeDomainEnd{1};

        \def\cInitial{1};
        \def\cBoundry{1.2};

        \begin{axis}[spaceTimeAxis]
            \draw[thick] (axis cs:0.1,0.5) -- (axis cs:0.9,0.5);

            \draw[thick,-{latex},greenSet, shorten <= 4pt] (axis cs:0.5,0.5) -- node [below right] {$p$} (axis cs:0.1,0.9);
            \draw[thick,-{latex},blueSet, shorten <= 4pt] (axis cs:0.5,0.5) -- node [below left] {$p$} (axis cs:0.9,0.9);

            \draw[thick,-{latex},blueSet, shorten >= 4pt] (axis cs:0.1,0.1) -- node [above right] {$p$} (axis cs:0.5,0.5);
                
            \node [] at (axis cs:0.9,0.4) {$\rho_1,c_1$};
            \node [] at (axis cs:0.9,0.6) {$\rho_2,c_2$};

        \end{axis}
    \end{tikzpicture}
 \caption{rightgoing}
 \end{subfigure}
  \caption{Time-scattering.}
  \label{fig:time_scattering}
\end{figure}
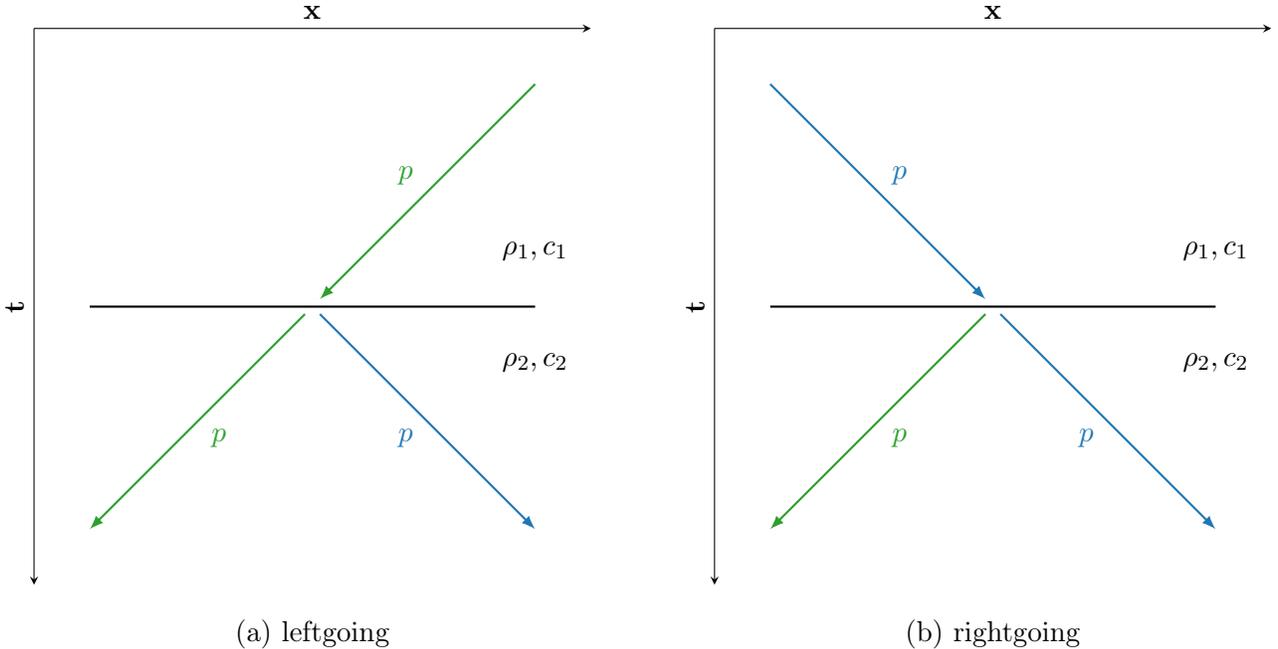

Scattering at a time boundary is schematically illustrated in Fig.~\ref{fig:time_scattering} for a left- and a right-going wave. The instant that the material properties change, the left- and right-going waves are partially reflected and transmitted with coefficients, $R^t$ and $T^t$, respectively.  

By transposing the space and time axes of Figs.~\ref{fig:time_scattering}, one notices a similarity with Fig.~\ref{fig:space_scattering}(right): due to causality, waves scattering at a time boundary cannot reflect into the past, just like we enforce ``spatial causality'' when constructing a focusing wavefield recursively by evaluating Eq.~\ref{eq:abovefrombelow2} in one direction only. This explains the absence of upward pointing arrows in Figs.~\ref{fig:time_scattering} and of rightward reflections in Figs.~\ref{fig:space_scattering}(right) and~\ref{fig:focusing_function}. Indeed, it has been noted that scattering at multiple time boundaries is simpler than scattering at multiple space boundaries because the infinite recursion due to waves reverberating between the space boundaries is avoided \citep{salem2015spacetime}. 

\begin{figure}
\setlength{\figW}{0.8\columnwidth}
\setlength{\figH}{\figW}
 
 \begin{tikzpicture}[tikzPlot]
    \def\timeBoundaryStart{0.4};
    \def\timeBoundaryEnd{0.6};
    \def\spaceBoundaryStart{0.5};
    
    \def\spaceDomainEnd{1};
    \def\timeDomainEnd{1};
    
    \def\cInitial{1};
    \def\cBoundry{1.2};
    
    \begin{axis}[spaceTimeAxis]
        
        \addplot[black,quiver={u=\thisrow{dx},v=\thisrow{dt}},thick,-] table [x={x}, y={t}]{timeBoundaries.dat};
        
        \addplot[blueSet,quiver={u=\thisrow{dx},v=\thisrow{dt}},thick,-{stealth[sep= 2pt]}] table [x={x}, y={t}]{raysRight.dat};
        \addplot[greenSet,quiver={u=\thisrow{dx},v=\thisrow{dt}},thick,-{stealth[sep= 2pt]}] table [x={x}, y={t}]{raysLeft.dat};
        
    \end{axis}
\end{tikzpicture}
  \caption{Wave scattering at multiple time boundaries.}
  \label{fig:focusing_function_time}
\end{figure}
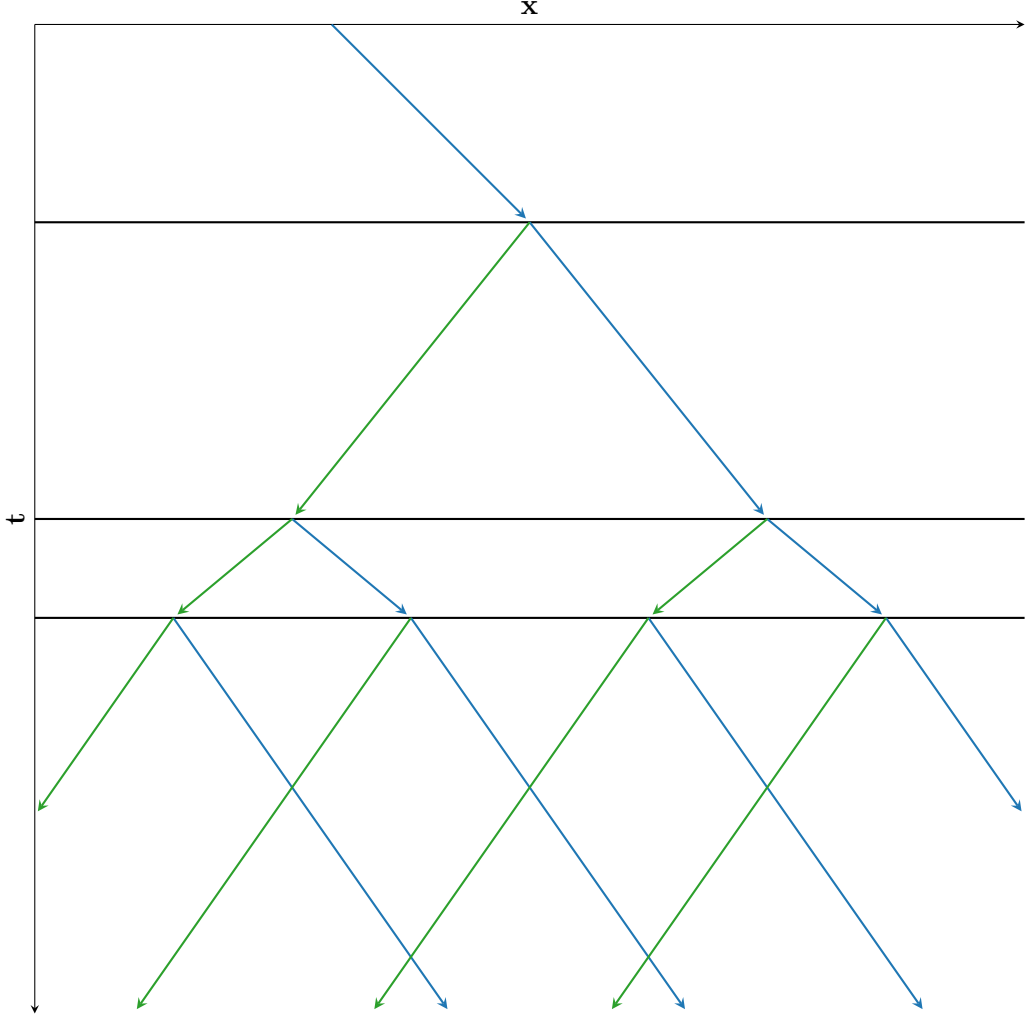

In Fig.~\ref{fig:focusing_function_time}, a wave scattering at multiple time boundaries is considered. At every time boundary, the wavefront splits into two, thus doubling the number of events. This situation can be compared to Fig.~\ref{fig:focusing_function}, for the focusing wavefield. In both figures, left-going waves are shown in green, while right-going waves are shown in dark blue. The similarities are striking, and one figure can be obtained from the other by transposing it \footnote{Note that even if the time-scattered and focusing wavefields can be mapped onto each other bijectively, the arrows are not all pointing in the same direction. Therefore, one must be cautious when interpreting this mapping and the corresponding physics.}. Now, the reason for the suggestive naming in Eqs.~\ref{eq:abovefrombelow2} is clear: In order to map the time-scattering onto the reverse space-scattering, the time transmission and reflection coefficients must have exactly the same \textit{form} as Eqs.~\ref{eq:inversecoeff}, as indeed we have shown.

\subsection*{Time and space delays and advances}
To finalize the mapping between the focusing and time-scattered wavefields, the relation between the layer thicknesses, $h^{x}_{j}$, and timing of the $N=2^{J+1}$ events in the focusing wavefield, $t^{d_1,d_2, \ldots, d_{J}}$, must be understood. Here, $J$ is the number of layers between the first and last interface, and $d_1, d_2, \ldots, d_{J}$ is a notation that will be explained shortly. Similarly, the relation between the layer duration, $h^{t}_{j}$, and the spacing of the events in the time-scattered field, $x^{d_1, d_2, \ldots, d_{J}}$, must be established. 

Starting with the focusing wavefield and taking the time the wave is transmitted into the rightmost layer as time zero, it is clear from Fig.~\ref{fig:focusing_function} that the timing of individual incident and reflected events can be found by adding or subtracting one-way times, $\Delta^{t}_{j} = h^{x}_{j} / c_{j}$, depending on the propagation direction of a wave segment in a particular layer. Denoting right-going waves by $-1$ and left-going waves by $+1$, and numbering layers and events in the negative space direction, we obtain:
\begin{equation}
t^{d_1,d_2,\ldots,d_{J}} = \sum_{j=1}^{J} d_j \Delta^{t}_{n} = \sum_{j=1}^{J} d_j \left( h^{x}_{j} / c_{j} \right) ,
\end{equation}
where $d_j$ is $\pm 1$ depending on the direction in layer $j$. 

Similarly, taking the position of the single, incident event as space zero, it is clear from Fig.~\ref{fig:focusing_function_time} that the spacing of individual transmitted and reflected events can be found by adding or subtracting one-way distances, $\Delta^{x}_{j} = c_{j} h^{t}_{j}$, depending on the propagation direction of a wave segment in a particular layer. Denoting right-going waves by $+1$ and left-going waves by $-1$, and numbering layers and events in the positive time direction, we obtain:
\begin{equation}
x^{d'_1,d'_2,\ldots,d'_{J}} = \sum_{j=1}^{J} d'_{j} \Delta^{x}_{n} = \sum_{j=1}^{J} d'_{j} \left( c_{j} h^{t}_{j} \right) ,
\end{equation}
where $d'_j$ is $\pm 1$ depending on the direction in layer $j$. 

\subsection*{Media that compute their own inverse}
The possibility of constructing a medium that computes its own inverse now becomes apparent when considering Fig.~\ref{fig:focusing_function_time} in combination with Fig.~\ref{fig:focusing_function}. Placing Fig.~\ref{fig:focusing_function_time} above and to the left of Fig.~\ref{fig:focusing_function} suggests that a single wave scattering at a sequence of time boundaries could yield the focusing wavefield that focuses through a sequence of space boundaries. 

Since the mapping between focusing wavefields and time-scattering is one-to-one, we only need to consider how a single space layer with thickness $h^{x}_{j}$ and propagation velocity $c_{j}$ maps onto a single time layer. In order to reproduce a particular one-way time delay or advance $\Delta^{t}_{j}$, we need to translate this time-delay to a one-way space delay or advance $\Delta^{x}_{j}$, respectively. Considering a connecting layer of constant propagation velocity, $c_c$, we can define the $\Delta^{x}_{j}$ in terms of the $\Delta^{t}_{j}$ as: $\Delta^{x}_{j} = c_c \Delta^{t}_{j}$. Hence, the time duration of the individual time layers can be found as:
\begin{equation}
h^{t}_{j} = \Delta^{x}_{j} / c_{j} = c_c \Delta^{t}_{j} / c_{j} = c_{c} h^{x}_{j} / c^{2}_{j}.
\label{eq:thickness2duration}
\end{equation}
Note that the time duration of the individual time layers, $h^{t}_{j}$, is indeed the only thing that needs to be computed since we have already established that the time  reflection and transmission coefficients feature the same velocities and densities that occur in the reverse space reflection and transmission coefficients. The time durations are a function of the propagation velocity of the connecting layer since the connecting layer is what allows mapping the time delays and advances in the focusing wavefield onto the space delays and advances in the time scattering. 

Finally, it must be mentioned here that while any connecting layer propagation velocity will yield a valid focusing wavefield that yields a single pulse on the other side of the spatial inhomogeneity, only a connecting layer propagation velocity equal to the propagation velocity of the first time layer / last spatial layer will yield a transmission coefficient of unity since only then the product of the individual frequency shifts occurring at every time boundary is unity. The same requirement appears in the next section when considering different boundary conditions.      

\subsection*{Validity of the construction for different boundary conditions}
Before illustrating the construction with a numerical example, we briefly show that the outlined construction remains valid when different boundary conditions (equivalent to taking $\mathbf{D}$ and $\mathbf{B}$ continuous in the electromagnetic case) between the time media are considered. In the case of rapid, continuously varying changes in properties, the system in Eqs.~\ref{eq:system_piecewise} is not a valid starting point. In Appendix~\ref{sec:appendixa} we consider a different system that is valid and derive the corresponding time transmission and reflection coefficients, $\tilde{T}^t$ and $\tilde{R}^t$, respectively as well as their relation with $T^t$ and $R^t$. Specifically, we find:
\begin{subequations}
\begin{eqnarray}
    \tilde{T}^t_{12} & = & \frac{1}{2}\left( \frac{c_2}{c_1}+\frac{\kappa_1}{\kappa_2} \right) = \left( \frac{c_2}{c_1} \right) T^t_{12} , \label{eq:Tt_continuously_compared} \\
    \tilde{R}^t_{12} & = & -\frac{1}{2}\left( \frac{c_2}{c_1}-\frac{\kappa_1}{\kappa_2} \right) = -\left( \frac{c_2}{c_1} \right) R^t_{12}. \label{eq:Rt_continuously_compared}
\end{eqnarray}
\label{eq:timecoeff_continuously_compared}
\end{subequations}
As we can see, the transmission and reflection coefficient for time media with the modified boundary conditions are scaled and inverted versions of the transmission and reflection coefficients for time media with field-continuous boundary conditions. 

At first sight, it seems that the additional minus sign appearing in the reflection coefficient destroys the one-to-one relation between the time scattered wavefields and the focusing wavefields since it changes (inverts) the ratio between the time transmission and reflection coefficient. However, since all the right-scattered waves in the time scattered wavefield undergo an even number of time-reflections (for an original right-going incident pulse), this is not the case. Furthermore, since the scaling does not affect the ratio of the time transmission and reflection coefficients, the resulting time scattered wavefield will still be a focusing wavefield. 

Second, it appears that the scaling may change the amplitude of the focusing wavefield. This is indeed the case. However, note that the final scale factor for a combination of time boundaries is the ratio of the last and the first layer velocities (i.e., after and before the ``stack'', respectively) since the intermediate velocities multiply out. Hence, provided that we consider a ``stack'' of time layers embedded in a time layer of constant velocities, the resulting scale factor will be unity and the resulting forward-scattered wavefield for the two types of time boundary conditions exactly the same.

\begin{figure*}
\includegraphics[trim=2.5cm 1.5cm 2.5cm 1.5cm,clip,width=0.85\linewidth]{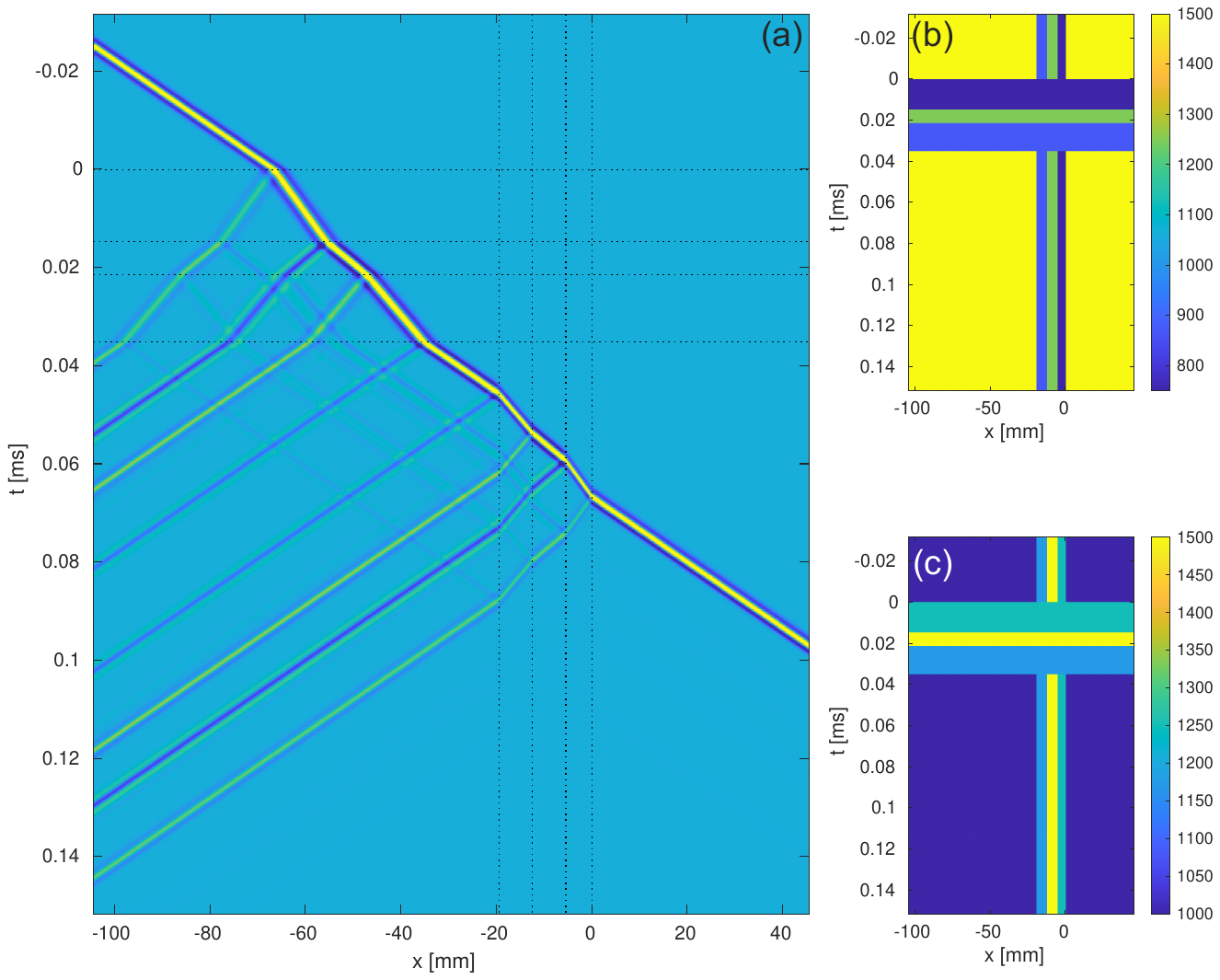}
\caption{Space-time medium that computes its own inverse. (a) Wavefield. (b) Propagation velocity [$m/s$]. (c) Density [$kg/m$].}
\label{fig:spacetime}
\end{figure*}

\section*{Results}

In Fig.~\ref{fig:spacetime}, a numerical example of a space-time medium that computes its own inverse is shown. In the right half of the medium, a spatial inhomogeneity is located, consisting of three layers of varying thickness, propagation velocity and density, embedded in a medium of higher propagation velocity (see Table~\ref{table:spaceprops}). Using Eq.~\ref{eq:thickness2duration} with Table~\ref{table:spaceprops} and a connecting layer propagation velocity $c_c = 1500~m/s$, the time durations of the corresponding temporal inhomogeneity were computed. The resulting material properties to be used in the first half of the simulation are summarized in Table~\ref{table:timeprops}. 

To model the propagation in the space-time medium, a staggered-grid finite-difference time-domain (FDTD) algorithm with simple absorbing boundaries was used~\footnote{To use simple absorbing boundaries, the simulation must be run at the maximum Courant number of unity, for which, `magically', there is no numerical dispersion in 1D. Note that this also explains why only lower velocities than the embedding propagation velocity of $c0 = 1500~m/s$ were used: any higher velocities would imply a Courant number smaller than unity for the embedding, resulting in boundary reflections.}. The material property matrices (propagation velocity and density) used in the simulation are shown in Fig.~\ref{fig:spacetime}(b) and~(c), respectively. For every time-step of the simulation, a corresponding row of these matrices is used. The resulting wavefield for an ultrasonic source on the left of the medium excited at early times with a Ricker wavelet ($2_{nd}$ derivative of a Gaussian) with center frequency $f_c = 300~kHz$ is shown in Fig.~\ref{fig:spacetime}(a). 

As can be seen in Fig.~\ref{fig:spacetime}(a), the single incident pulse first scatters into two, then four, then eight, and finally into sixteen events upon encountering the four time boundaries (present after $t=0$). Eight of these events are right-going and propagate towards the spatial inhomogeneities (present before $x=0$). Upon interacting with the spatial inhomogeneity, we can see that the timing of these incident waves is such that rightward reflections of leftward reflections of the initial pulse are suppressed. Thus, only a single transmitted pulse is observed on the right of the spatial inhomogeneity confirming that the field incident on the spatial inhomogeneity indeed consists of the focusing wavefield. Furthermore, inspection of the wavelet and amplitude of the single transmitted pulse confirms that it has the same wavelet and amplitude as the original incident pulse, such that we can say that the transmission coefficient of the inhomogeneous space-time medium is unity and therefore that the medium has computed its own inverse.     

The complete Matlab code used to compute the example is included in Appendix~\ref{sec:appendixb}.

\begin{table}
\centering
\begin{tabular}{||p{1cm} p{1.5cm} p{2cm} p{1.5cm}||} 
 \hline
 Layer & Thickness [mm] & Propagation velocity [m/s] & Density [kg/m] \\ [0.5ex] 
 \hline\hline
 1 & $\infty$ & 1500 & 1000 \\ 
 \hline
 2 & 7 & 875 & 1175 \\
 \hline
 3 & 7 & 1250 & 1500 \\
 \hline
 4 & 5.5 & 750 & 1250 \\
 \hline
 5 & $\infty$ & 1500 & 1000 \\  
 \hline
\end{tabular}
\caption{Space medium properties.}
\label{table:spaceprops}

\vspace*{0.5 cm}

\centering 
\begin{tabular}{||p{1cm} p{1.5cm} p{2cm} p{1.5cm}||} 
 \hline
 Layer & Duration [ms] & Propagation velocity [m/s] & Density [kg/m] \\ [0.5ex] 
 \hline\hline
 1 & $\infty$ & 1500 & 1000 \\ 
 \hline
 2 & 0.01371 & 875 & 1175 \\
 \hline
 3 & 0.00672 & 1250 & 1500 \\
 \hline
 4 & 0.01467 & 750 & 1250 \\
 \hline
 5 & $\infty$ & 1500 & 1000 \\  
 \hline
\end{tabular}
\caption{Time medium properties ($c_c=1500~m/s$).}
\label{table:timeprops}

\end{table}

\section*{Discussion and Conclusion}
The existence of an acoustic space-time medium that computes its own inverse has been demonstrated. Because the medium is time- and space-variant, basic properties such as linear time invariance (LTI) and existence of an impulse response and the output being the convolution of an arbitrary input with such an impulse response are no longer guaranteed. However, we note that the construction of the medium does not rely on a fixed space-time `offset' of the time and space boundaries: provided that the connecting layer is `large' enough (in time and space), an impulse response can be defined and longer inputs can be considered for which the output will still be the convolution with the impulse response. Conversely, note that even for a single incident pulse, there must be sufficient space available to accommodate all the time scattering and in the example both the extent of the domain and the timing of the first time boundary have been chosen such that the scattering at the last time boundary takes place before the leftmost left-going wave hits the left edge of the domain and the rightmost right-going wave hits the spatial discontinuity. Since it is only the relative spacing/timing of the incident waves that matters, provided that we are able to extend the background/embedding medium, it is always possible to create enough space-time between the time boundaries and the space boundaries such that a medium computing its own inverse can be constructed.

When the connecting layer has the same propagation velocity as the first time and last space layer, the transmission coefficient of the system is unity. However, the left-reflected field is clearly non-zero, consisting of both left-going waves resulting from the time scattering as well as leftward reflections of the focusing wavefield. 

The possible applications of such a medium are many. One could think of environments where it is physically impossible to place hardware, but where the medium as a whole can be manipulated. In such a case, a single incoming pulse can be manipulated such that it anticipates certain undesired reverberations at spatial inhomogeneities. The most likely candidate for experimental realization is ultrasonics in fluids, where previous work has shown the ability of modifying the acoustic medium through electro-rheological effects~\cite{Li_2017}. We are just beginning to explore the possibilities of wave control using space-time boundaries.

\begin{acknowledgments}
This work received funding from SNF grant 197182.
\end{acknowledgments}

\appendix
\section{Time transmission and reflection coefficients for continuously varying properties}
\label{sec:appendixa}
Eqs.~\ref{eq:system_piecewise} are valid for time media with piece-wise constant material properties. However, if the time medium is continuous, that is the boundaries represent rapid continuously varying changes in medium properties, system~\ref{eq:system_piecewise} is not valid and the starting point should instead be:
\begin{subequations}
\begin{eqnarray}
    \partial_x p + \partial_t (\rho v_x) & = & f_x, \\
    \partial_x v_x + \partial_t (\kappa p) & = & q.
\end{eqnarray}
\label{eq:system_continuously}
\end{subequations}
In this case, the time derivatives act on quantities $\rho v_x$ and $\kappa p$ and hence those quantities should be taken continuous at a time boundary. Defining quantities $\Theta = -\kappa p$ and $m_x = \rho v_x$ the boundary conditions can be written:
\begin{subequations}
\begin{eqnarray}
    \Theta(x, t=0^{-}) & = & \Theta(x, t=0^{+}), \\
    m_x(x, t=0^{-}) & = & m_x(x, t=0^{+}),
\end{eqnarray}
\label{eq:cc_time_continously}
\end{subequations}
With these definitions, the incident field in Eq.~\ref{eq:pre_time} can be rewritten:
\begin{equation}
\begin{rcases}
    \Theta = & -\kappa_1 p = -\kappa_1 p_i\delta(x - c_1 t) \\
    m_x = & \rho_1 v_x = c_1^{-1} p_i\delta(x - c_1 t) 
\end{rcases}
\text{ $t < 0 $ }.
\label{eq:pre_time_continuously}
\end{equation}    
And the transmitted and reflected field in Eq.~\ref{eq:post_time} can be rewritten:
\begin{equation}
\begin{rcases}
    \Theta = & -\kappa_2 p = -\kappa_2 \left[ p_t\delta(x - c_2 t) + p_r\delta(x + c_2 t) \right] \\
    m_x = & \rho_2 v_x = c_2^{-1}\left[p_t\delta(x - c_2 t) - p_r\delta(x + c_2 t)\right] 
\end{rcases}
\text{ $t > 0 $ }.
\label{eq:post_time_continuously}
\end{equation}    
Using the new continuity conditions, \ref{eq:cc_time_continously}, in combination with Eqs.~ \ref{eq:pre_time_continuously} and~\ref{eq:post_time_continuously} above, we obtain:
\begin{subequations}
\begin{eqnarray}
    -\kappa_1 p_i & = & -\kappa_2 \left( p_t + p_r \right) , \\
    c_1^{-1} p_i & = & c_2^{-1} \left( p_t - p_r \right) .
\end{eqnarray}
\end{subequations}
Solving the system of equations for $p_t$ and $p_r$ in terms of $p_i$ and defining the new transmission and reflection coefficients $\tilde{T}^t=p_t/p_i$ and $\tilde{R}^t=p_r/p_i$, we find:
\begin{subequations}
\begin{eqnarray}
    \tilde{T}^t_{12} & = & \frac{1}{2}\left( \frac{c_2}{c_1}+\frac{\kappa_1}{\kappa_2} \right) , \label{eq:Tt_continuously} \\
    \tilde{R}^t_{12} & = & -\frac{1}{2}\left( \frac{c_2}{c_1}-\frac{\kappa_1}{\kappa_2} \right) , \label{eq:Rt_continuously}
\end{eqnarray}
\label{eq:timecoeff_continuously}
\end{subequations}
where, as before, the first digit in the subscript of the transmission and reflection coefficient defines the incident medium. 

Comparing Eqs.~\ref{eq:timecoeff_continuously} to Eqs.~\ref{eq:timecoeff}, we see that the coefficients obtained when taking the equivalent of $\mathbf{D}$ and $\mathbf{B}$ continuous are related to the coefficients obtained when taking the equivalent of $\mathbf{E}$ and $\mathbf{H}$ continuous as follows:
\begin{subequations}
\begin{eqnarray}
    \tilde{T}^t_{12} & = & \left( \frac{c_2}{c_1} \right) T^t_{12} , \label{eq:Tt_compared} \\
    \tilde{R}^t_{12} & = & -\left( \frac{c_2}{c_1} \right) R^t_{12} . \label{eq:Rt_compared}
\end{eqnarray}
\label{eq:timecoeff_compared}
\end{subequations}

\section{Matlab code for the example}
\label{sec:appendixb}

\begin{lstlisting}[
  mathescape = true,
  style      = Matlab-editor,
  basicstyle = \mlttfamily,
]
r0=1000; c0=1500; scf=10000;
dx=1/scf; nx=1500; xs=dx*(0:nx-1);
dt=dx/c0; nt=2750; ts=dt*(0:nt-1);
ix=5; fc=30*scf; tsrc=2/fc;
rm=r0*ones(nt,nx);
cm=c0*ones(nt,nx);

ilayerx0=850;  
ilayerx=ilayerx0;
hlayerx=[70,70,55];  
clayerx=[875,1250,750];
rlayerx=[1175,1500,1250]; 
nlayerx=length(hlayerx);
for ii=1:nlayerx
  cm(:,ilayerx+(1:hlayerx(ii)))=clayerx(ii);
  rm(:,ilayerx+(1:hlayerx(ii)))=rlayerx(ii);
  ilayerx=ilayerx+hlayerx(ii);
end

tlayerx=dx*hlayerx./clayerx; 
tlayert=c0*dx*hlayerx./clayerx.^2;

ilayert0=475;  
ilayert=ilayert0;
hlayert=round(tlayert/dt);
clayert=clayerx;
rlayert=rlayerx;
nlayert=nlayerx;
hlayert=fliplr(hlayert);
clayert=fliplr(clayert);
rlayert=fliplr(rlayert);
for ii=1:nlayert
  cm(ilayert+(1:hlayert(ii)),:)=clayert(ii);
  rm(ilayert+(1:hlayert(ii)),:)=rlayert(ii);
  ilayert=ilayert+hlayert(ii);
end

[p_all,vx_all]=fdtd1_abc(rm,...
  cm,dt,nt,dx,nx,ix,fc,tsrc);

figure; imagesc(1000*xs,1000*ts,p_all); 
xlabel('x [mm]'); ylabel('t [ms]');
caxis(caxis/2);

%--------------------------------------------

function [p_all,vx_all]=fdtd1_abc(rm,...
  cm,dt,nt,dx,nx,ix,fc,tsrc)

p_all=zeros(nt,nx); p=zeros(1,nx);
vx_all=zeros(nt,nx); vx=zeros(1,nx);
c1m=dt.*rm.*cm.^2; c2m=dt./rm; 
ts=dt*(0:nt-1);
term1=(1-2*pi^2*fc^2*(ts-tsrc).^2);
term2=exp(-(pi*fc*(ts-tsrc)).^2);
qsrc=term1.*term2; 
        
for a=1:nt
  vx(1,nx)=vx(1,nx-1); % simple abc
  vx(1,1:nx-1)=vx(1,1:nx-1)-...
    c2m(a,1:nx-1).*diff(p,1,2)./dx;
  p(1,1)=p(1,2); % simple abc
  p(1,2:nx)=p(1,2:nx)-...
    c1m(a,2:nx).*diff(vx,1,2)/dx;   
  p(1,ix)=p(1,ix)+qsrc(a);
  p_all(a,:)=p;
  vx_all(a,:)=vx;    
end  

\end{lstlisting}


\bibliography{apssamp.bib}

\end{document}